\documentclass[a4paper,10pt,twoside]{cpc-hepnp}

\usepackage{multicol}
\usepackage{graphicx}
\usepackage{booktabs}
\usepackage{amssymb,bm,mathrsfs,bbm,amscd}
\usepackage[tbtags]{amsmath}
\usepackage{lastpage}

\begin{document}

\fancyhead[c]{\small Chinese Physics C~~~Vol. 37, No. 1 (2013) 010201}
\fancyfoot[C]{\small 010201-\thepage}

\footnotetext[0]{Received 14 March 2009}

\title{Rotational properties in even-even superheavy $^{254-258}$Rf nuclei based on total-Routhian-surface calculations\thanks{Supported by National Natural Science
Foundation of China (10805040,11175217), Foundation and Advanced
Technology Research Program of Henan Province(132300410125), S \& T
Research Key Program of Henan Province Education Department
(13A140667). }}

\author{%
      WANG Hua-Lei$^{1;1)}$\email{wanghualei@zzu.edu.cn}%
\quad CHAI Qing-Zhen$^{1}$ \quad JIANG Jin-Ge$^{1}$\quad LIU
Min-Liang$^{2}$ } \maketitle

\address{%
$^1$  School of Physics and Engineering, Zhengzhou University,
Zhengzhou 450001, China\\
$^2$ Institute of Modern Physics, Chinese Academy of Sciences,
Lanzhou 730000, China\\
}

\begin{abstract}
High-spin yrast structures of even-even superheavy nuclei
$^{254-258}$Rf are investigated by means of total-Routhian-surface
approach in three-dimensional ($\beta_2, \gamma, \beta_4$) space.
The behavior in the moments of inertia of $^{256}$Rf is well
reproduced by our calculations, which is attributed to the
$j_{15/2}$ neutron rotation-alignment. The competition between
rotationally aligned $i_{13/2}$ proton and $j_{15/2}$ neutron is
discussed. High-spin predictions are also made for its neighboring
isotopes $^{254,258}$Rf.
\end{abstract}

\begin{keyword}
rotational properties, total-Routhian-surface calculation, moment of
inertia, band crossing
\end{keyword}

\begin{pacs}
21.60.Ev, 
21.10.Re, 
27.90.+b, 

\end{pacs}

\footnotetext[0]{\hspace*{-3mm}\raisebox{0.3ex}{$\scriptstyle\copyright$}2013
Chinese Physical Society and the Institute of High Energy Physics
of the Chinese Academy of Sciences and the Institute
of Modern Physics of the Chinese Academy of Sciences and IOP Publishing Ltd}%
\begin{multicols}{2}

\section{Introduction}

Both experimentally and theoretically, studies of superheavy nuclei,
with $Z\gtrsim 104$, are a fast developing piece of nuclear
physics~\cite{Sobiczewski2007,Schadel2006}. These nuclei are
stabilized by a shell-correction energy, which originates from the
clustering of single-particle orbitals and the occurrence of regions
of low level density. The structures and formation mechanism of
shell-stabilized nuclei are governed by the similar physics
responsible for the predicted stability of an island of superheavy
nuclei around $Z=114, N=184$ (also $Z=120, 124,$ or 126, depending
on theoretical models and the parametrization
employed)~\cite{Nilsson1968,Rutz1997, Zhang2012}. The precise
location of the proposed new magic numbers depends on the
single-particle structure sensitively. Presently, the related
single-particle levels of the new spherical shell cannot be directly
investigated through the use of both in-beam and decay spectroscopy
because the production cross sections of these nuclei with $Z\geq
110$ are very small.

However, a great number of spectroscopy measurements have been
carried out to study yrast and non-yrast level structures, including
high-K isomers, for the deformed nuclei of the transfermium mass
region, typically with $Z\approx 100, N\approx
150-160$~\cite{Herzberg2004, Greenlees2007, Eeckhaudt2005}.
High-spin rotational bands have been observed in many of them (e.g.
$^{247,249}_{96}$Cm~\cite{Tandel2010},
$^{248,249,250,252}_{98}$Cf~\cite{Tandel2010, Takahashi2010},
$^{250}_{100}$Fm~\cite{Bastin2006},
$^{251}_{101}$Md~\cite{Chatillon2007},
$^{252,253,254}_{102}$No~\cite{Reiter1999,Reiter2005,Herzberg2009,Herzberg2001},
$^{255}_{103}$Lr~\cite{Ketelhut2009}). Theoretically, the nuclear
properties of the transfermium nuclei have been systematically
investigated using various theoretical methods such as the
macroscopic-microscopic models and the self-consistent mean-field
models (see Refs.~\cite{Rutz1997,Zhang2012, Liu2012, Zheng2009,
AlKhudair2009} and references therein). Though these nuclei are not
really superheavy ones, they are at the gateway to the superheavy
mass region. In particular, their high-spin information can provide
an indirect way to study the single-particle states of the next
spherical shell above $^{208}_{82}$Rf$_{126}$ because some of them
are strongly down-sloping and thus can come close to the Fermi
surface in the deformed region~\cite{Chen2008}. Moreover, these
high-spin states also may give information on the fission barrier at
high angular momentum which is important for understanding the
produced mechanism for the heaviest elements~\cite{Reiter1999}.

For the superheavy isotopes, spectroscopic studies are in a
preliminary stage and detailed information on high-spin properties
is still scarce~\cite{Berryman2011}. Nevertheless, the ground-state
band in the $Z=104$ isotope $^{256}_{104}$Rf$_{152}$, the first
even-even superheavy element for which high-spin data now exist, was
recently reported by Greenlees {\it et al}~\cite{Greenlees2012}.
Soon afterwards, the kinematic and dynamic moments of inertia of the
ground-state band in $^{256}_{104}$Rf$_{152}$ was studied by using a
particle-number conserving method based on a cranked shell
model~\cite{Zhang2013}. It is conceivable that with the help of the
new generation detector systems and radioactive beam
facilities~\cite{Wang2010,Liu2011}, the experimental spectroscopic
borders of this region will soon be expanded towards even heavier
nuclei. In the present work, we aim to understand the high-spin
states in the nucleus $^{256}$Rf using self-consistent
total-Routhian-surface (TRS) approach based on Woods-Saxon (WS)
potential and predict the rotational behaviors of its neighboring
even-even isotopes $^{254,258}$Rf. Part of the aim of this work is
to test the predictive power of present model in superheavy nuclei,
although generally successful in other mass regions. The different
polarization effects and functional forms of the densities may occur
in superheavy region. These effects can be naturally incorporated
within the self-consistent nuclear mean-field models, but the
preconceived knowledge about the expected densities and
single-particle potentials, which fades away towards superheavy
region, is required in the macroscopic-microscopic
investigations~\cite{Rutz1997}.

\section{The model}

The TRS approach~\cite{Nazarewicz87,Satula94npa,Xu00}, which is
based on the cranked shell model~\cite{Bengtsson79,Frauendorf81},
accounts well for the overall systematics of high-spin phenomena in
rapidly rotating medium and heavy mass nuclei. The total Routhian,
which is called "Routhian" rather than "energy" in a rotating frame
of reference, is the sum of the energy of the non-rotating state and
the contribution due to cranking. The energy of the non-rotating
state consists of a macroscopic part that is obtained from the
standard liquid-drop model~\cite{Myers66} and a microscopic term
representing the Strutinsky shell correction~\cite{Strutinsky67}.
Single-particle energies needed in the calculation of the quantal
shell correction are obtained from a triaxial WS
potential~\cite{Nazarewicz85, Cwiok87} with the parameter set widely
used for cranking calculations. During the diagonalization process
of the WS hamiltonian, the oscillator states with the principal
quantum number $N \leqslant 12$ and 14 have been used as a basis for
protons and neutrons, respectively. The nuclear shape is defined by
the standard parametrization in which it is expanded in spherical
harmonics~\cite{Cwiok87}. The deformation parameters include
$\beta_2, \gamma$, and $\beta_4$ where $\gamma$ describes nonaxial
shapes.  The pairing correlation is treated using the Lipkin-Nogami
approach~\cite{Pradhan73} in which the particle number is conserved
approximately. This avoids the spurious pairing phase transition
encountered in the simpler BCS calculation. Not only monopole but
also doubly stretched quadrupole pairings are considered. The
quadrupole pairing is important for the proper description of the
moments of inertia, though it has negligible effect on
energies~\cite{Satula95}. The monopole pairing strength, $G$, is
determined by the average gap method \cite{Moller92} and the
quadrupole pairing strengths are obtained by restoring the Galilean
invariance broken by the seniority pairing
force~\cite{Xu00,Sakamoto90}. The 80 single-particle levels, the
respective 40 states just below and above the Fermi energy, are
included in the pairing windows for both protons and neutrons. The
Strutinsky quantal shell correction is performed with a smoothing
range $\gamma =1.20\hbar \omega _0$, where $\hbar \omega
_0=41/A^{1/3}$ MeV, and a correction polynomial of order $p=6$.

The nuclear system is cranked around a fixed axis (the x-axis) at a
given rotational frequency $\omega$. Pairing correlations are
dependent on rotational frequency and deformation. The resulting
cranked-Lipkin-Nogami equation takes the form of the well known
Hartree-Fock-Bogolyubov-like (HFB) equation~\cite{Xu00}. For a given
rotational frequency and point of deformation lattice, pairing is
treated self-consistently by solving this equation using a
sufficiently large space of WS single-particle states. Certainly,
symmetries of the rotating potential can be used to simplify the
cranking equations. In the reflection-symmetric case, both
signature, $r$, and intrinsic parity, $\pi$ are good quantum
numbers. The solution characterized by ($\pi$, $r$) provides
simultaneously the energy eigenvalue from which it is
straightforward to obtain the energy relative to the non-rotating
state. After the numerical calculated Routhians at fixed $\omega$
are interpolated using cubic spline function between the lattice
points, the equilibrium deformation can be determined by minimizing
the calculated TRS. The absolute minimum of the TRS corresponds to
the solution for an yrast state. Secondary ones correspond to other
solutions, which may be yrast at higher spins.

\section{Results and discussions}

\begin{center}
\includegraphics[width=8cm]{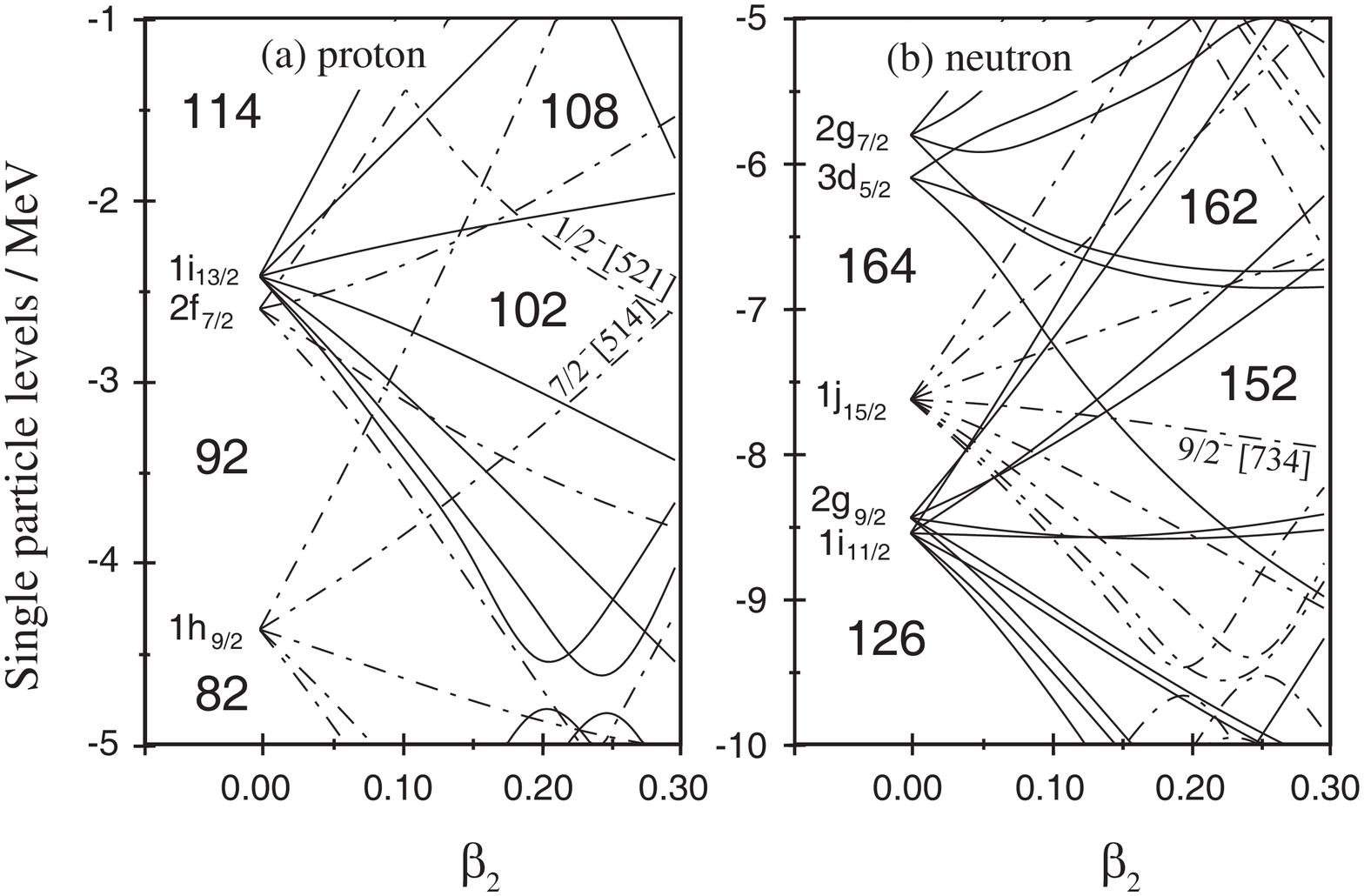}\\
\figcaption{\label{spe}  Calculated Woods-Saxon single particle
levels near the Fermi surface of $^{256}_{104}$Rf$_{152}$ (a) for
protons and (b) for neutrons. The positive (negative) parity levels
are denoted by solid (dot-dash) lines.}
\end{center}

To test the validity of current nuclear potential, single-particle
levels calculated for $^{256}_{104}$Rf$_{152}$ are shown in
Fig.~\ref{spe}, for protons and neutrons, respectively. The typical
spherical and deformed shell gaps can clearly be seen, which
indicate the best candidates for spherical and deformed nuclei. As
an example, the combination of the $Z=108$ and $N=162$ gaps at
$\beta_2\approx0.24$ leads to $^{270}_{108}$Hs$_{162}$, which was
predicted to be doubly magic deformed nucleus~\cite{Patyk1991}. The
effect of the known deformed shell at the neutron number N=152 is
also reproduced in the results for the mass region. The properties
of the single-particle structures are in good agreement with
previous study by Sobiczewski {\it et al}~\cite{Sobiczewski2001}.
The present results are also supported by the microscopic
corrections calculated by M\"{o}ller {\it et al}~\cite{Moller1995},
which are related to single-particle level density (Low level
density results in the large correction energy). In Fig.~\ref{spe},
the unique-parity high-$j$ intruder orbitals are $1i_{13/2}$ proton
and $1j_{15/2}$ neutron orbitals. The proton (neutron) Fermi surface
is just above $\pi 1/2^-[521]$ ($\nu 9/2^-[734]$) orbital, which is
filled in $Z=103$ and 104 ($N= 151$ and 152) nuclei. As known, the
study of odd-$A$ nuclei can provide important information on
single-particle structures. In present calculation, the sequence of
single-proton levels near the Fermi surface is quite similar with
that determined from the experimental information of
$^{255}_{103}$Lr$_{152}$ isotone~\cite{Chatillon2006}, in which the
energies of $1/2^-[521](2f_{5/2})$ and $7/2^-[514](1h_{9/2})$ are
nearly degenerate. In $^{255}_{104}$Rf$_{151}$
isotope~\cite{Hebberger2006}, the ground-state configuration is
suggested to be $9/2^-[734](1j_{15/2})$ by experiment, which is also
consistent with our calculation.

\begin{center}
\includegraphics[width=8cm]{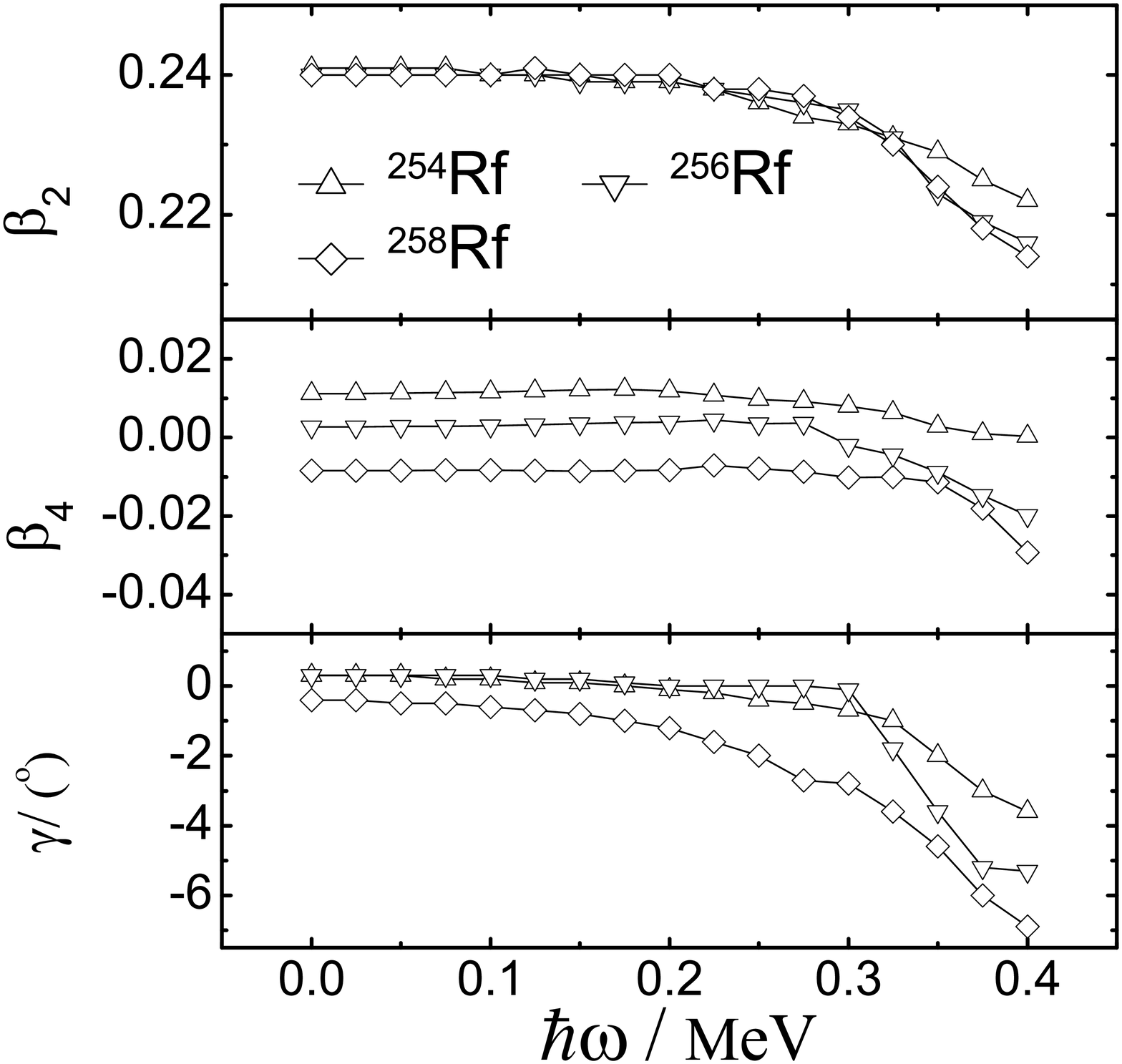}\\
\figcaption{\label{Def} Calculated deformations $\beta_2$ (top),
$\beta_4$ (middle) and $\gamma$ (bottom) vs the rotational frequency
$\hbar \omega$ for even-even nuclei $^{254-258}$Rf.  }
\end{center}

In principle, the study of multi-quasiparticle excitations can be
used to determine the excitation energies and ordering of the
single-particle states near the Fermi surface. For example, two
2-quasiproton isomeric states are reported in $^{254}$No by Herzberg
{\it et al}~\cite{Herzberg2006}, one with a K-value and parity of
$K^\pi = 8^-$($9/2^+[624] \otimes 7/2^-[514]$) and the other with a
$K^\pi = 3^+$($1/2^-[521] \otimes 7/2^-[514]$). The excitation
energies of the $3^+$ and $8^-$ states were found at 0.988 and 1.296
MeV, which indicate the single-particle energy of $1/2^-[521]$
orbital is lower than that of $9/2^+[624]$ orbital. Similarly,
2-quasiparticle isomeric states in $^{256}$Rf are expected to be
identified, which can provide a test for current nuclear model.
Experiments on quasiparticle excitations for $^{256}$Rf have been
performed by Jeppesen {\it et al}~\cite{Jeppesen2009} and by
Robinson {\it et al}~\cite{Robinson2011}. However, the search for
2-quasiparticle states is not very successful, even their
experimental results and interpretation are in disagreement and
conflict. Up to now, there is no convincing evidence for the
existence of 2-quasiparticle states in $^{256}$Rf, additional
experimental investigation is thus highly desirable.

The present TRS calculations based on the WS potential are performed
in the lattice of quadrupole ($\beta_2, \gamma$) deformations with
hexadecapole ($\beta_4$) variation, since our previous
investigation~\cite{Wang2012} with the inclusion of
reflection-asymmetric deformation degrees of freedom shows that
there is no octupole instabilities at normal deformed minima for the
lighter Rf isotopes. In Fig.~\ref{Def}, the evolutions of calculated
$\beta_2$, $\beta_4$ and $\gamma$ deformations with increasing
rotational frequency are displayed for even-even $^{254-258}$Rf
nuclei studied here. These nuclei are predicted to have
well-deformed prolate shapes with very small $\gamma$ deformation.
Note that the calculated ground-state $\beta_2$ deformations,
typically $\approx 0.24$, are basically consistent with the previous
results given by M\"{o}ller {\it et al}~\cite{Moller1995} and by
Sobiczewski {\it et al}~\cite{Sobiczewski2001}. The associated
$\beta_2$ deformations are rather stable and decrease slightly with
increasing rotational frequency. In our opinion, the decrease of
$\beta_2$ as a function of rotational frequency is due to the mixing
between the $g$-band and a 2-quasiparticle ($\pi i_{13/2}$ or $\nu
j_{15/2}$ quasi-particles) aligned band. It seems that a rapid
decrease of $\beta_2$ appears after a critical frequency $\approx
0.3$ MeV, which indicates the band crossing frequency. For prolate
deformed nuclei, the rotation-induced deoccupation of the
antialigned high-$j$ low-$\Omega$ quasiparticle orbitals can cause
the nucleus to become less deformed~\cite{Petrache1998}. The $\pi
i_{13/2}$ alignment and the $\nu j_{15/2}$ alignment seem to produce
slightly triaxial shapes.

\begin{center}
\includegraphics[width=8cm]{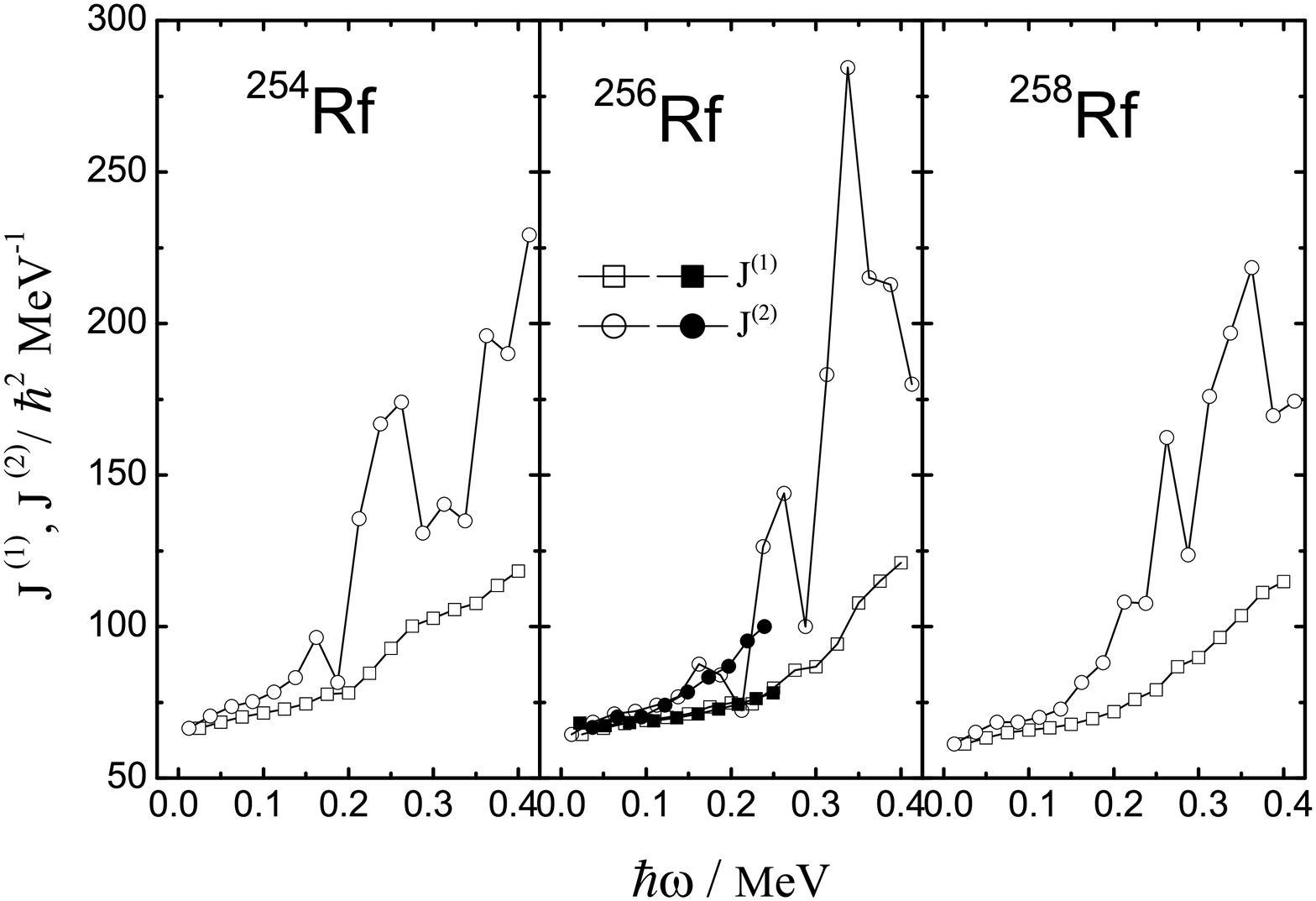}\\
\figcaption{\label{J12} Kinematic and dynamic moments of inertia
$J^{(1)}$ and $J^{(2)}$ of even-even nuclei $^{254-258}$Rf vs the
rotational frequency $\hbar \omega$ from experiments (filled
symbols) and present calculations (open symbols). Experimental data
are taken from Ref.~\cite{Greenlees2012}. }
\end{center}

Figure~\ref{J12} shows the kinematic and dynamic moments of inertia,
$J^{(1)}$ ($J^{(1)}$ = $\hbar^2(2I-1)/E_{\gamma}(I)$) and $J^{(2)}$
($J^{(2)}$ = $4\hbar^2/[E_{\gamma}(I)-E_{\gamma}(I-2)]$), for the
ground-state bands of even-even nuclei $^{254-258}$Rf, which usually
can provide useful information on the energies of single-particle
orbitals, particularly of those with high $j$. The level spins
adopted here are obtained from Ref.~\cite{Greenlees2012}, which are
further confirmed by Zhang {\it et al}~\cite{Zhang2013}. One can see
that the agreements of the moments of inertia for $^{256}$Rf between
theory and experiment are to a great extent excellent. The moments
of inertia $J^{(1)}$, $J^{(2)}$ show a gradual increase, reflecting
a large-interaction band crossing, and essentially no difference
among even-even $^{254-258}$Rf nuclei, especially at low frequency.
The upbending phenomena due to the alignment of $j_{15/2}$ neutrons
and $i_{13/2}$ protons are clearly seen in Fig.~\ref{J12}. In
general, the changes in the moments of inertia can be well
understood in terms of the changing deformations or/and pairing
correlations. The quadrupole deformation $\beta_2$ usually has an
important effect on the moments of inertia. However, from the
$\omega$ dependences of $\beta_2$ in Fig.~\ref{Def}, we observe a
nearly constant behavior for the low-spin region and a sudden
decrease in the high-spin states where rotation-alignment occurs.
Obviously, the increase in moment of inertia, as shown in
Fig.~\ref{J12}, cannot be explained by an increasing deformation
$\beta_2$. It should be attributed to a decrease in the nucleon
pairing correlations due to the Coriolis anti-pairing (CAP) effect.
The slightly negative $\gamma$ deformation may result in a change in
the moment of inertia.

\begin{center}
\includegraphics[width=8cm]{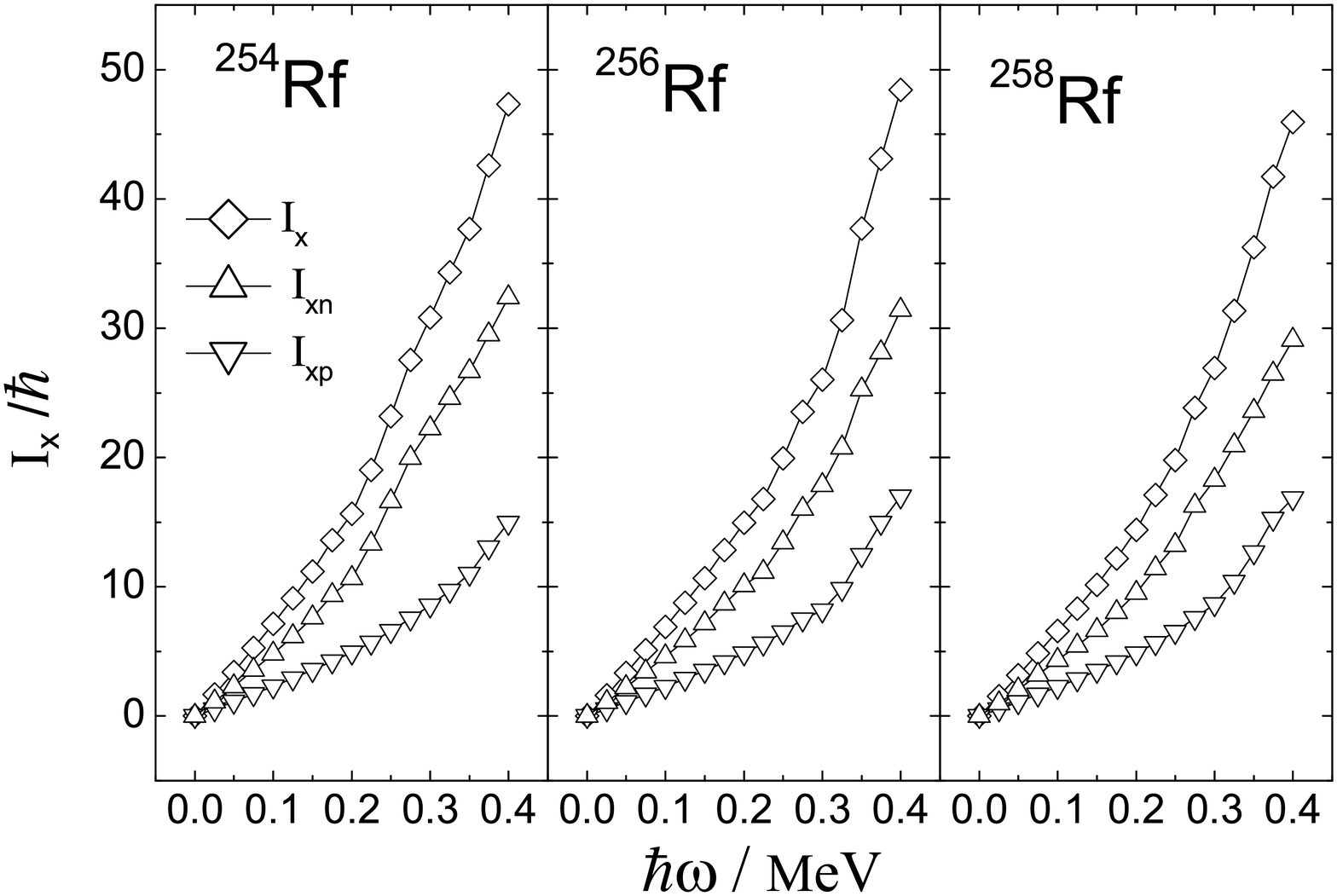}\\
\figcaption{\label{Ix} Calculated aligned angular momentum as a
function of rotational frequency $\hbar \omega$ for even-even nuclei
$^{254-258}$Rf. Proton and neutron contributions are shown
simultaneously. }
\end{center}

To understand the changing properties in the moments of inertia, we
display the calculated collective angular momenta for even-even
$^{254-258}$Rf nuclei in Fig.~\ref{Ix}, together with proton and
neutron components. As expected, the aligned angular momentum shows
a similar changing trend to the moment of inertia (Note that its
slope means the moment of inertia $J^{(2)}$). It can be seen that a
substantial increase of the total aligned angular momentum is due to
the neutron contribution, though both proton and neutron components
increase gently with increasing frequency. A sudden rise in the
aligned angular momentum indicates additional contribution from the
rotation-alignment of a pair of high-$j$ nucleons. In this mass
region, the related high-$j$ single-particle states are proton
$i_{13/2}$ and neutron $j_{15/2}$ orbitals. About 30 years ago, the
alignment properties of these orbitals in the actinides were
surveyed by Frauendorf {\it et al}~\cite{Frauendorf81,Chen1983} and
by Simon {\it et al}~\cite{Simon1980,Simon1982}. Moreover, the
behaviour of the simultaneous alignment of protons and neutrons has
been noticed in different nuclei such as in $^{235}$U and
$^{237}$Np~\cite{Frauendorf81}. Recently, several
studies~\cite{Zhang2012, AlKhudair2009, Afanasjev2003} indicated the
alignments of $i_{13/2}$ proton and $j_{15/2}$ neutron pairs are
also strongly competitive in some transfermium nuclei (heavy
actinides). For example, it was pointed out that the proton and
neutron alignments take place simultaneously in the lighter No
isotopes (e.g. in $^{252,254}$No), while the $j_{15/2}$ neutrons win
in the competition between proton- and neutron-alignment in the
heavier No isotopes (e.g. in $^{256,258}$No)~\cite{AlKhudair2009}.
Figure~\ref{Ix} shows that the first band crossing is most likely to
be ascribed to rotation-alignment of a pair of $j_{15/2}$ neutrons
for the lighter $^{254}$Rf nucleus and there is a possible
competition in rotation-alignment between high-j intruder protons
and neutrons for $^{256,258}$Rf, because the proton and neutron
upbending frequencies seem to be somewhat close to each other.

\begin{center}
\includegraphics[width=8cm]{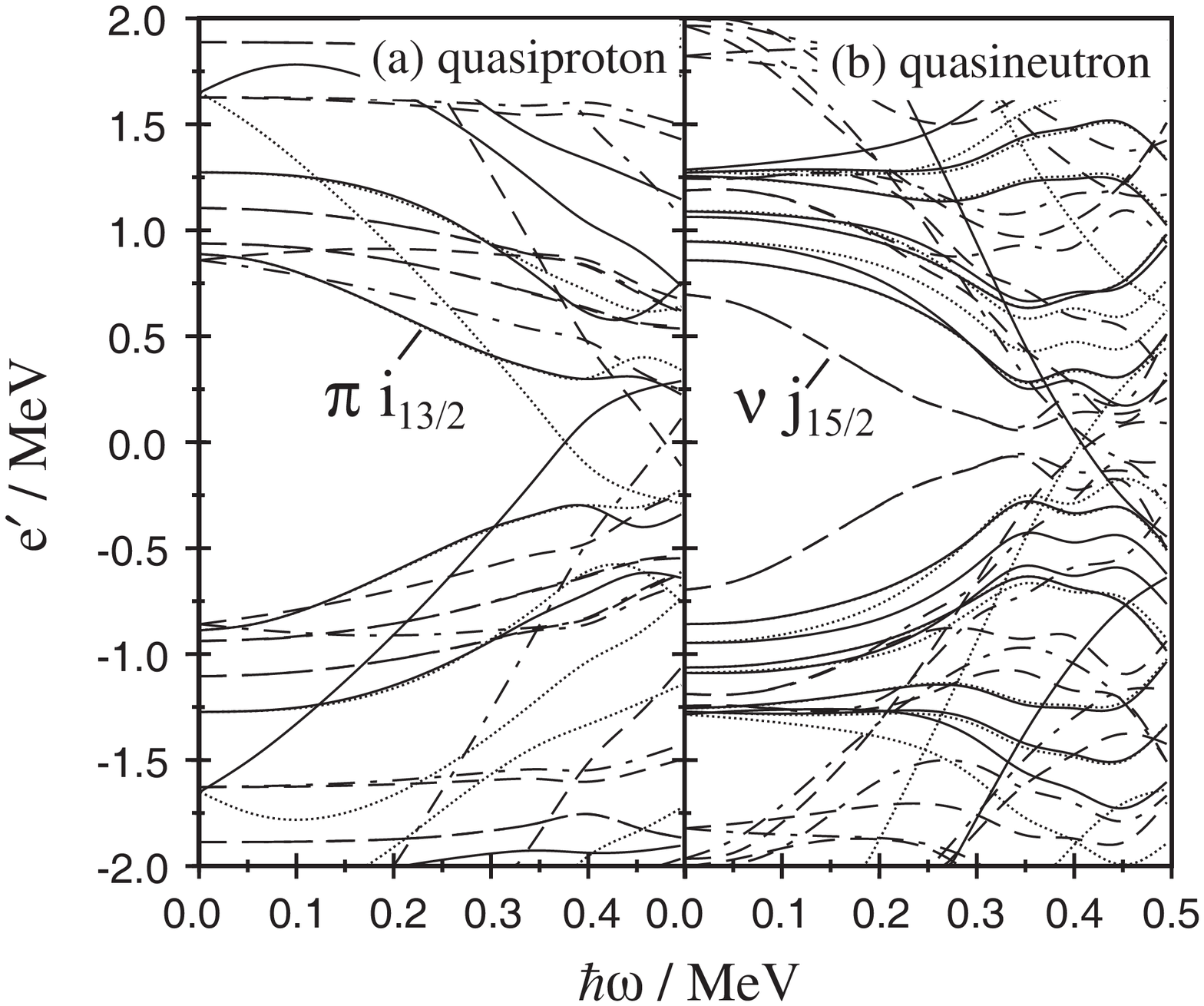}\\
\figcaption{\label{qpg} Calculated quasiproton Routhians (a) and
quasineutron Routhians (b) as functions of rotational frequency
$\hbar \omega$. The deformation parameters used correspond to the
yrast configurations of even-even nucleus $^{256}$Rf. The parity and
signature ($\pi, \alpha$) of the Routhians are presented as follows:
($+, +1/2$) solid lines, ($+, -1/2$) dotted lines, ($-, +1/2$)
dashed-dotted lines, and ($-, -1/2$) dashed lines.}
\end{center}

The diagram of the quasi-particle routhians versus rotational
frequency is a convenient way to display the interplay between
rotation and microscopic structure. By studying the quasi-particle
diagram, as shown in Fig.~\ref{qpg}, one can see the relevant
quasi-particles are those from the unique-parity high-j intruder
orbitals with $\pi i_{13/2}$ for protons and $\nu j_{15/2}$ for
neutrons. The $\nu j_{15/2}$ band crossing in $^{256}$Rf is
calculated to occur slightly earlier than that for $\pi i_{13/2}$,
which indicates a competitive behavior to some extent. In contrast,
our calculation predicts that the alignment of the $1j_{15/2}$
neutron pair is more favored than the $\pi i_{13/2}$ alignment in
$^{254, 258}$Rf. Moreover, similar to the situation in $N= 152$
$^{254}$No isotone~\cite{Liu2012}, a delayed alignment is calculated
to occur in $^{256}$Rf.

\section{Summary}

In the present work, we have studied the evolution properties of the
deformations and moments of inertia for even-even $^{254-258}$Rf
nuclei by using self-consistent TRS method based on WS potential in
($\beta_2, \gamma, \beta_4$) deformation space. Our results are
compared with previous calculations and available experiments. It is
shown that the single-particle potential is still valid in the
superheavy region and the deformed shell gaps (e.g. at $N=152$ and
162) are well reproduced. Calculated moments of inertia of
$^{256}$Rf are in good agreement with experiments and the up-bending
is attributed to rotation-alignment of $1j_{15/2}$ neutron pair,
showing a competitive property to some extent. Our prediction of
rotational properties for $^{254, 258}$Rf awaits future experimental
confirmation. It is clear, however, that high-spin experiments on
$^{254, 258}$Rf, whose correction energies are smaller than that of
the deformed magic number $N= 152$ $^{256}$Rf nucleus, will meet a
great challenge due to the large fission probabilities.
\\

%
\end{multicols}

\vspace{15mm}

\centerline{\rule{80mm}{0.1pt}}
\vspace{2mm}

\begin{multicols}{2}

\end{multicols}

\clearpage

\end{document}